\documentclass[structabstract]{aa}  
\usepackage{graphicx}
\usepackage{txfonts}
\usepackage[authoryear]{natbib}
\begin{document}
   \title{Characterization of Kepler early-type targets\thanks{Based on observations collected with the
         telescope at the {\it M.G. Fracastoro} station of the INAF - Osservatorio Astrofisico di Catania.}}

   \author{G. Catanzaro
          \inst{1}
          \and
          A. Frasca\inst{1}
          \and
          J. Molenda-\.Zakowicz\inst{2}
          \and
          E. Marilli\inst{1}
          }

   \institute{INAF - Osservatorio Astrofisico di Catania,
              Via S. Sofia 78, 95123 Catania, Italy\\
             \email{gca@oact.inaf.it}
             \and
             Astronomical Institute of the University of Wroc\l aw, 
             ul.\ Kopernika 11, 51-622. Wroc\l aw, Poland
             }

   \date{Received , 2010; accepted , 2010}

 
  \abstract
   {Stellar pulsation offers a unique opportunity to constrain the intrinsic parameters 
    of stars and unveil their inner structure. The Kepler satellite is collecting an enormous 
    amount of data of unprecedent photometric precision, which will allow us to test theory
    and obtain a very precise tomography of stellar interiors.}
   {We attempt to determine the stars' fundamental parameters (T$_{\rm eff}$, $\log g$, 
    v$\sin i$, and luminosity) needed for computing asteroseismic models and 
    interpreting Kepler data. We report spectroscopic observations of 23 early-type Kepler 
    asteroseismic targets, 13 other stars in the Kepler field, that had not been selected to be 
    observed.}
   {We measured the radial velocity by performing a cross-correlation with template spectra
    to help us identify non-single stars.
    Spectral synthesis was performed to derive the stellar parameters of our target stars, and 
    the state-of-the-art LTE atmospheric models were computed. For all the stars of our sample, 
    we derived the radial velocity, T$_{\rm eff}$, $\log g$, 
    v$\sin i$, and luminosities. For 12 stars, we performed a detailed abundance 
    analysis of 20 species, for 16, we could derive only the [Fe/H] ratio. 
    A spectral classification was also performed for 17 stars in the sample.}
   {We identify two double-lined spectroscopic binaries, HIP\,96299 and HIP\,98551, the
    former of which is an already known eclipsing binary,
    and two single-lined spectroscopic binaries, HIP\,97254 and HIP\,97724.
    We also report  two suspected spectroscopic binaries, HIP\,92637 and HIP\,96762,  
    and the detection of a possible variability in the radial velocity of HIP\,96277. 
    Two of our program stars are chemically peculiar, namely HIP\,93941, 
    which we classify as B2\,He-weak, and HIP\,96210, which we classify as B6\,Mn.
    Finally, we find that HIP\,93522, HIP\,93941, HIP\,93943, HIP\,96210 and HIP\,96762, 
    are very slow rotators ($v\sin i < 20$ km\,s$^{-1}$) which makes them very 
    interesting and promising targets for asteroseismic modeling.}
   {}


   \maketitle
%

\section{Introduction}
Kepler is a NASA space mission that was successfully launched on 6 March 2009. Its primary goal 
is the detection of Earth-size and larger planets orbiting around stars similar to the Sun by means 
of the method of photometric transits \citep[see][]{borucki2009}. 
Kepler photometry will also be used to detect pulsations in stars selected as Kepler asteroseismic
targets by the Kepler Asteroseismic Science Consortium, KASC\footnote{http://astro.phys.au.dk/KASC}.
The study of their pulsational properties will allow us to investigate their internal structure and 
to derive accurate stellar radii by means of asteroseismic methods \citep[see][]{jcd2007}.
The first results of the Kepler mission, published by \citet{borucki09}, confirm the expected high 
precision of the acquired data, which allows to discover Earth-size planets on Earth-like orbit around 
stars as faint as 12 mag.

In reality, most of the atmospheric parameters, such as T$_{\rm eff}$, $\log g$, and metallicity needed
by theoreticians to compute asteroseismic models, come from the Kepler Input Catalogue (KIC), which 
is based on calibrated photometry. Therefore, any determination of these atmospheric parameters based on 
ground-based spectroscopic data of Kepler targets is valuable and important. In this paper, we present 
a detailed spectroscopic analysis of 34 stars of spectral type earlier than F selected 
in 2007 from the list of candidate Kepler targets with Hipparcos parallaxes provided by \citet{molenda2006}. 
At that time, the Kepler mission was still in preparation and the exact position of the Kepler field 
of view had not been fixed.
As a result, it was not known which of our targets would be finally observed by Kepler.
Thus we decided to acquire spectra of all the stars with Hipparcos data that were potential asteroseismic 
targets in the sky region where Kepler was to have been pointed. 
After defining the mission,  65\% of the stars proposed by us were selected as
Kepler asteroseismic targets by the steering committee of the KASC.
Some results concerning cool stars were published by \citet{molenda07,molenda08}.

For each star, we derive the parameters needed to construct asteroseismic models, i.e., the effective 
temperature, surface gravity, and the metal abundance. Dealing with asteroseismic radius determination,
which means combining stellar atmospheric parameters with large frequency spacing, \citet{stello2009}
show that an accuracy in T$_{\rm eff}$ of $\approx$\,200~K is enough to guarantee an error in the
radius of smaller than 4$\%$ for stars as hot as $\approx$\,6400~K. The same authors also show that the most
important parameter is metallicity, an accuracy of $\sigma\,[Fe/H] \approx\,$0.15 dex corresponding
to an error in the radius on the order of 2$\%$. To achieve this accuracy, moderate- or high-resolution
spectroscopy is fully adequate.

Projected rotational velocity ($v\sin i$) is another essential parameter for constraining asteroseismic
models. It provides information about the possible multiplet splitting caused by rapid rotation that 
could change the structure of the frequency spectrum. In this sense, stars with low values of $v\sin i$ 
are the most promising asteroseismic targets for Kepler.

Finally, for all our program stars we measure the radial velocity, $RV$. This can be used to detect new
spectroscopic binaries for which we may need to correct the stars' Kepler magnitudes, as well as
the magnitudes and colors obtained form ground-based data, for the presence of the secondary component. 
These corrections are important when deriving the stars' effective temperature and luminosity from photometry. 
Treating a binary system as a single star may indeed result in the computation of an incorrect asteroseismic 
model, a spurious interpretation of the frequency spectrum, and a wrong determination of the radius of the star 
and of any eventual planet. The presence of an unknown component can also influence the frequency pattern of 
the star with its own frequencies and make our interpretation of the power spectrum totally wrong.

\section{Observation and data reduction}

\subsection{Photometry}

The photometric observations were carried out in the Str\"omgren {\it uvby}$\beta$ system 
with the 91-cm Cassegrain telescope at {\it M.G. Fracastoro} station (Serra La Nave, Mt. Etna, 1735\,m a.s.l.) 
of the {\it INAF\,-\,Osservatorio Astrofisico di Catania} (OAC).
The observations were performed with a photon-counting photometer equipped with an EMI 9893QA/350 
photomultiplier, cooled to $-15\,\degr$. Owing to poor weather and technical problems, we could observe only 
14 of our targets along with three stars in the same field used to determine the zero
points (see Table\,\ref{tab:standubvy}). The observations were performed on 2008 
December 1 and 2009 June 6 and 7.

A set of {\it uvby} standard stars of known Str\"omgren indices \citep{Lin73}
and $V$ magnitude were nightly observed to transform the instrumental 
magnitudes  of the targets into the Str\"omgren photometric system. Both the 
$V$ magnitude and Str\"omgren photometric indices of all the target stars that we 
were able to observe are given in Table~\ref{tab:programubvy}.

\begin{table}
\caption[ ]{Magnitudes and colors of the zero-point stars from \citet{HauckMerm}.}
\label{tab:standubvy}
\begin{tabular}{lccccc} 
\hline 
\hline 
\noalign{\smallskip}
 HD      &   $V$   &   $b-y$ &  $m_{1}$  &  $c_{1}$ &  $\beta$\\ 
\noalign{\smallskip}
\hline 
\noalign{\smallskip}
 188209  &   5.65  &   0.009 &  0.033    & $-0.107$ & 2.549 \\    
 189013  &   6.89  &   0.089 &  0.176    & 1.015    & 2.843 \\  
 215399  &   8.71  &   0.283 &  0.135    & 0.516    & 2.669 \\ 
\noalign{\smallskip} 
\hline
\end{tabular}
\end{table}

The average errors are 0.010, 0.010, 0.020, and 0.020 mag for $V$, $b-y$, m$_{1}$, and c$_{1}$,
respectively. For the $beta$ index, we find a typical error of 0.020.	

\begin{table*}
\caption[ ]{$V$ magnitude and Str\"omgren parameters for a subset of the program stars.}
\label{tab:programubvy}
\begin{tabular}{lrrrrrrrrr} 
\hline 
\hline 
\noalign{\smallskip}
 HIP      &   $V$~~~~~~   &   $b-y$~~~~ &  $m_{1}$~~~~~~  &  $c_{1}$~~~~~~~ &  $\beta$~~~~~~~  &  $T_{\rm eff}$~ & $\log g$ & $M_V$ & $\Delta m_0$ \\ 
\noalign{\smallskip}
\hline 
\noalign{\smallskip}
91178 &  9.161\,$\pm$\,0.010  & $-0.029\,\pm$\,0.010  &  0.214\,$\pm$\,0.010  &  1.122\,$\pm$\,0.010  & 2.872\,$\pm$\,0.020  &  9720  & 3.69 & $-0.02$  & $-0.083$  \\  
92247 &  9.307\,$\pm$\,0.010  &  0.111\,$\pm$\,0.010  &  0.239\,$\pm$\,0.020  &  0.840\,$\pm$\,0.030  & 2.786\,$\pm$\,0.020  &  7510  & 3.87 &  2.01   & $-0.042$  \\  
92259 &  9.524\,$\pm$\,0.010  &  0.133\,$\pm$\,0.010  &  0.182\,$\pm$\,0.010  &  1.067\,$\pm$\,0.020  & 2.813\,$\pm$\,0.020  &  7770  & 3.46 &  0.42   &   0.012   \\  
92637 &  9.840\,$\pm$\,0.010  & $-0.017\,\pm$\,0.010  &  0.039\,$\pm$\,0.020  &  0.167\,$\pm$\,0.010  & 2.598\,$\pm$\,0.020  & 17980  & 2.85 & $-3.50$  &   0.021   \\  
93070 &  9.410\,$\pm$\,0.010  &  0.070\,$\pm$\,0.010  &  0.187\,$\pm$\,0.025  &  1.069\,$\pm$\,0.025  & 2.825\,$\pm$\,0.020  &  7860  & 3.52 &  0.45   &   0.016   \\  
93522 & 10.196\,$\pm$\,0.010  &  0.029\,$\pm$\,0.015  &  0.204\,$\pm$\,0.025  &  1.065\,$\pm$\,0.010  & 2.883\,$\pm$\,0.020  &  8760  & 3.93 &  0.96   & $-0.022$  \\  
93924 &  9.139\,$\pm$\,0.010  & $-0.021\,\pm$\,0.010  &  0.168\,$\pm$\,0.015  &  0.958\,$\pm$\,0.030  & 2.873\,$\pm$\,0.030  & 10280  & 4.17 &  0.77   & $-0.028$  \\ 
93941 & 10.656\,$\pm$\,0.020  & $-0.081\,\pm$\,0.020  &  0.105\,$\pm$\,0.030  & $-0.054\,\pm$\,0.020  & 2.699\,$\pm$\,0.020  & 28360  & 5.15 & $-2.23$  & $-0.047$  \\  
93943$^1$ &  9.730\,$\pm$\,0.010  &  0.137\,$\pm$\,0.015  &  0.174\,$\pm$\,0.025  &  1.043\,$\pm$\,0.020  & 2.814\,$\pm$\,0.020  &  7770  & 3.53 & 0.65  &   0.019   \\  
94137 &  9.849\,$\pm$\,0.010  &  0.186\,$\pm$\,0.010  &  0.179\,$\pm$\,0.040  &  0.813\,$\pm$\,0.030  & 2.810\,$\pm$\,0.020  &  7810  & 4.24 & 2.70   &   0.006   \\  
95495 &  9.268\,$\pm$\,0.010  &  0.089\,$\pm$\,0.010  &  0.292\,$\pm$\,0.010  &  0.831\,$\pm$\,0.010  & 2.810\,$\pm$\,0.020  &  7800  & 4.14 & 2.45   & $-0.090$  \\  
95506 &  9.218\,$\pm$\,0.020  &  0.001\,$\pm$\,0.020  &  0.181\,$\pm$\,0.020  &  1.111\,$\pm$\,0.020  & 2.873\,$\pm$\,0.020  &  9300  & 3.72 & 0.26   & $-0.020$  \\  
96066$^2$ &  8.579\,$\pm$\,0.020  &  0.041\,$\pm$\,0.020  &  0.182\,$\pm$\,0.020  &  0.937\,$\pm$\,0.020  & 2.916\,$\pm$\,0.020  &  9040  & 4.48 & 2.16   &   0.001   \\ 
96776 &  9.371\,$\pm$\,0.020  &  0.209\,$\pm$\,0.020  &  0.068\,$\pm$\,0.020  &  1.123\,$\pm$\,0.020  & 2.760\,$\pm$\,0.020  &  7290  & 2.92 & $-0.82$  &   0.100   \\
\noalign{\smallskip} 
\hline
\end{tabular} \\
\begin{flushleft}
$^1$ A fainter companion ($\Delta V=1\fm7$) at about 9$\arcsec$ was included in the diaphragm. \\
$^2$ Combined magnitude and colors. The fainter companion HIP\,96061 ($V=9\fm64$) at 11$\farcs$5 was 
also included in the diaphragm. \\
\end{flushleft}
\end{table*}

\subsection{Spectroscopy}
The spectra of our program stars were acquired on 13 nights between June 18 and September 17, 2007 
and on three nights from July to October 2009 (see Table~\ref{param}) at the {\it M.G. Fracastoro} 
station of OAC. We used the 91-cm telescope and FRESCO, the fiber-fed REOSC echelle spectrograph of 
OAC which allows us to obtain spectra in the range of 
4300--6800 {\AA} with a resolution R=\,21\,000. 
The spectra were recorded on a thinned, back-illuminated (SITE) CCD with 1024$\times$1024 pixels 
of 24~$\mu$m size, whose typical readout noise is about 8 e$^{-}$ and gain is 2.5 e$^-$/ADU.

The reduction of spectra, which included the subtraction of the bias frame, trimming, correcting
for the flat-field and the scattered light, the extraction of the orders, and the wavelength 
calibration, was performed using the NOAO/IRAF package\footnote{IRAF is
distributed by the National Optical Astronomy Observatory, which is operated by the
Association of Universities for Research in Astronomy, Inc.}. 
The amount of scattered light correction was about 10 ADU. 
After dividing the extracted spectra by a flat-field, the residual shape of the spectrum had been 
removed by dividing each spectral order by a Legendre function of a low order. Typical signal-to-noise 
ratio (SNR) of our spectra is $\sim$\,50. Finally, the IRAF package {\sf rvcorrect} was used to 
compute the velocity correction for the Earth's motion, which moved the spectra into 
the heliocentric rest frame.

The $V$ magnitudes of our targets range from about 9 to 11 mag since the sample was optimized for stars 
that are bright enough to have high signal-to-noise ratio in Kepler photometry but do not saturate Kepler 
CCDs. We acquired spectra for 23 stars falling onto the active pixels of Kepler CCDs, and for 13 
stars that had not previously benn observed by Kepler or that occupied the gaps on the CCDs or that were 
not selected as asteroseismic targets by the KASC.

\section{Radial velocity}
We measured the radial velocity, $RV$, of our program stars with the IRAF task {\sf fxcor}
by means of the method of cross-correlation \citep[see, e.g.,][]{Tonry79,Fitz93} between each order of 
the echelle spectra of the targets and the corresponding echelle order of the spectrum of HR\,1389 
(A2IV-V, $v_{\rm hel}=38.97\,\rm km\,s^{-1}$) or $\iota$\,Psc (F7\,V, $v_{\rm hel}=5.4\, \rm km\,s^{-1}$) \citep[see][]{Fekel99, Nord04} selected as $RV$ standard stars.  
We excluded from the analysis all spectral ranges heavily affected 
by telluric absorption lines (e.g., the O$_2$ band $\lambda$6276--6315). 
For a precise evaluation of the centroids of the CCF peaks in each echelle order, we fitted 
them with a Gaussian function. To deblend the two CCF peaks seen in SB2 systems, a 
two-Gaussian fit algorithm was applied.

The radial velocity of program stars were calculated by averaging $RV$ measurements from
all the echelle orders with the usual instrumental weight $W_{\rm i}=\sigma_{\rm i}^{-2}$.
The $\sigma_{\rm i}$ values were computed by {\sf fxcor}, which takes into account the height 
of the fitted peak and the antisymmetric noise as described by \citet{Tonry79}.
The standard errors in the weighted means were computed on the basis of the errors $\sigma_{\rm i}$ 
in the $RV$ values for each order \citep[see, e.g.,][]{Topp72}. 

We list the radial velocities of our program stars along with their uncertainties 
in Table~\ref{param}.

\begin{table*}[th]
\begin{minipage}{17.7cm}
\caption{Stellar parameters derived for the 23 Kepler asteroseismic targets in the KIC Catalogue and 13 other stars in 
Kepler field of view, but not yet selected as mission targets.} 
\label{param} 
\begin{center}                
\begin{tabular}{r r r r c r c r c r r}                       
\hline            
\hline            
HIP~     & KIC~~~~ &  $V^{\rm a}$~~  & Plx$^{\rm b}$~~~~ &Sp.     & $T_{\rm eff}$~~~~      & $\log g$   & $v\,\sin i$   & HJD & $RV$~~~~~~ & $\log L/L_\odot^{\rm c}$\\
         &         & (mag)& (mas)~~~~  &Type    & (K)~~~~           &            & (km/s) & +2450000    &  (km/s)~~~~ &            \\
\hline\noalign{\smallskip}
\multicolumn{11}{c}{\bf Kepler asteroseismic targets}\\
\noalign{\smallskip}\hline
92247    &     11013201 &  9.24 & 4.64\,$\pm$\,0.79   &	...     &  7200\,$\pm$\,200  & 3.5\,$\pm$\,0.2 & 100 & 4309.492 &  $-$8.3\,$\pm$\,2.6 & 0.87$_{-0.18}^{+0.16}$ \\
92259    &     10187831 &  9.47 & 3.29\,$\pm$\,0.77   & A7\,$\pm$\,3 &  7100\,$\pm$\,200  & 3.0\,$\pm$\,0.5 & 120 & 4310.465 & $-$22.7\,$\pm$\,2.6 & 1.11$_{-0.25}^{+0.20}$ \\
92637    &     10960750 &  9.83 & $-0.70\,\pm$\,0.85 & O9\,$\pm$\,2 & 19900\,$\pm$\,1400 & 3.8\,$\pm$\,0.2 & 230 & 4307.480 & $-$22.0\,$\pm$\,2.8 & 1.38$_{-0.40}^{+0.28}$  \\
93070    &      3830233 &  9.36 & 2.61\,$\pm$\,0.93   &	... 	      &  8000\,$\pm$\,500  & 4.2\,$\pm$\,0.2 & 200 & 4306.580 &    13.4\,$\pm$\,3.2 & ...~~~~~ \\
93522    &      8740371 & 10.14 & 0.12\,$\pm$\,1.     & A7\,$\pm$\,2 &  9100\,$\pm$\,400  & 3.7\,$\pm$\,0.2 &   5 & 4308.483 &     1.2\,$\pm$\,0.4 & ...~~~~~ \\
93941$^1$&      6848529 & 10.64 & 5.79\,$\pm$\,1.03   & B2$\pm$\,2   & 19300\,$\pm$\,1000 & 3.8\,$\pm$\,0.2 &  10 & 4270.370 & $-$14.3\,$\pm$\,3.0 & 0.85$_{-0.19}^{+0.16}$ \\
~~~$\prime\prime$~~~ &~~~$\prime\prime$~~~ &~~~$\prime\prime$~~~&~~~$\prime\prime$~~ &~~~$\prime\prime$~~   &~~~$\prime\prime$~~  &~~~$\prime\prime$~~~ & $\prime\prime$ & 5017.515 & $-$21.3\,$\pm$\,2.9 & ...~~~~~  \\
93943    &     12250891 &  9.70 &  5.26\,$\pm$\,1.58  &  ...              &  7400\,$\pm$\,200  & 3.4\,$\pm$\,0.5 &  20 & 4311.492 & $-$13.6\,$\pm$\,0.3 & 0.57$_{-0.33}^{+0.25}$ \\
94670    &      7599132 &  9.30 &  1.27\,$\pm$\,0.90  & B9\,$\pm$\,2 & 11600\,$\pm$\,500  & 4.2\,$\pm$\,0.2 &  50 & 4308.354 & $-$57.2\,$\pm$\,1.8 & ...~~~~~ \\
94809    &      6769635 &  9.32 &  3.54\,$\pm$\,0.72  &	... 	      &  7200\,$\pm$\,400  & 3.9\,$\pm$\,0.5 & 150 & 4290.393 &     5.8\,$\pm$\,4.4 & 1.10$_{-0.22}^{+0.18}$ \\
95092    &     11134456 &  9.83 &  2.62\,$\pm$\,0.87  &	... 	      &  8300\,$\pm$\,400  & 3.5\,$\pm$\,0.3 &  80 & 4307.546 & $-$11.2\,$\pm$\,2.1 & 1.19$_{-0.37}^{+0.27}$ \\
95174    &      5786771 &  9.08 &  1.91\,$\pm$\,0.76  & A2\,$\pm$\,3 & 10700\,$\pm$\,500  & 4.2\,$\pm$\,0.2 & 200 & 4306.520 & $-$21.7\,$\pm$\,12.1 & 1.97$_{-0.46}^{+0.31}$ \\
95495    &     11762256 &  9.23 &  3.91\,$\pm$\,0.75  &	... 	      &  7000\,$\pm$\,200  & 3.5\,$\pm$\,0.5 &  90 & 4308.542 & $-$43.9\,$\pm$\,1.8 & 1.04$_{-0.20}^{+0.17}$ \\
96210$^2$&     6128830  &  9.20 &  0.62\,$\pm$\,0.74  & B6\,$\pm$\,2 & 12600\,$\pm$\,600  & 3.5\,$\pm$\,0.3 &  15 & 4310.392 &     6.6\,$\pm$\,1.0 & \\
96277    &     10604429 &  9.93 &  2.70\,$\pm$\,1.11  &	... 	      &  7200\,$\pm$\,200  & 3.5\,$\pm$\,0.5 &  60 & 4278.437 &  $-$3.1\,$\pm$\,1.8 & 1.11$_{-0.48}^{+0.32}$ \\
~~~$\prime\prime$~~~ &~~~$\prime\prime$~~~ &~~~$\prime\prime$~~~&~~~$\prime\prime$~~ & &~~~$\prime\prime$~~~ &~~~$\prime\prime$~~~ &$\prime\prime$ & 4313.459 &     0.7\,$\pm$\,1.2 & ...~~~~~ \\
~~~$\prime\prime$~~~ &~~~$\prime\prime$~~~ &~~~$\prime\prime$~~~&~~~$\prime\prime$~~ & &~~~$\prime\prime$~~~ &~~~$\prime\prime$~~~ &$\prime\prime$ & 4361.440 &  $-$0.8\,$\pm$\,1.6 & ...~~~~~ \\
96299\,A &      3858884 &  9.25 &  2.49\,$\pm$\,1.06  &  ...             & ...~~~~       & ...                    &  70 & 4290.556 &  $-$7.8\,$\pm$\,1.2 & ...~~~~~ \\  
96299\,B &~~~$\prime\prime$~~~ &~~~$\prime\prime$~~~&~~~$\prime\prime$~~ & ... & ...~~~~       & ... &...         & 4290.556 &    45.1\,$\pm$\,1.3 & ...~~~~~ \\  
96762    &      4276892 & 9.13 &  2.71\,$\pm$\,0.76   & B9\,$\pm$\,1 & 10800\,$\pm$\,600  & 4.1\,$\pm$\,0.2 &  10 & 4309.399 &    16.5\,$\pm$\,1.1 & 1.61$_{-0.31}^{+0.23}$ \\
97486    &      9663677 & 9.98 &  2.51\,$\pm$\,0.87   &	... 	      &  8000\,$\pm$\,300  & 3.9\,$\pm$\,0.4 & 160 & 4312.533 &  $-$0.8\,$\pm$\,4.8 & 1.17$_{-0.39}^{+0.28}$  \\
97582    &      7978223 & 9.22 &  4.12\,$\pm$\,0.86   &	... 	      &  7100\,$\pm$\,200  & 3.6\,$\pm$\,0.5 &  90 & 4270.494 &     3.6\,$\pm$\,3.5 & 0.99$_{-0.22}^{+0.18}$ \\
~~~$\prime\prime$~~~ &~~~$\prime\prime$~~~&~~~$\prime\prime$~~~ &~~~$\prime\prime$~~~& &~~~$\prime\prime$~~~ &~~~$\prime\prime$~~~ &$\prime\prime$ & 4361.495 &	7.1\,$\pm$\,6.7 & ...~~~~~ \\
~~~$\prime\prime$~~~ &~~~$\prime\prime$~~~&~~~$\prime\prime$~~~ &~~~$\prime\prime$~~~& &~~~$\prime\prime$~~~ &~~~$\prime\prime$~~~ &$\prime\prime$ & 4312.593 &	2.6\,$\pm$\,2.9 & ...~~~~~ \\
97254    &      4581434 &  9.07 &  1.65\,$\pm$\,0.80  & A2\,$\pm$\,3 & 10200\,$\pm$\,200  & 4.2\,$\pm$\,0.2 & 200 & 4306.403 &  $-$4.3\,$\pm$\,6.2  & 2.08$_{-0.60}^{+0.36}$ \\
97724    &      4681323 &  9.05 &  1.85\,$\pm$\,0.76  &	... 	      &  8900\,$\pm$\,400  & 3.5\,$\pm$\,0.2 &  90 & 4308.590 &  $-$14.8\,$\pm$\,4.0 & 1.87$_{-0.48}^{+0.32}$ \\ 
98037    &      5304891 &  9.16 &  1.78\,$\pm$\,0.74  & B6\,$\pm$\,2 & 13100\,$\pm$\,700  & 3.9\,$\pm$\,0.2 & 180 & 4309.550 &  $-$30.9\,$\pm$\,20.4 & 2.21$_{-0.49}^{+0.32}$ \\
98160    &      8389948 &  9.16 &  2.79\,$\pm$\,0.75  &	... 	      & 10000\,$\pm$\,300  & 4.0\,$\pm$\,0.2 & 130 & 4307.418 &  $-$31.7\,$\pm$\,5.4 & 1.50$_{-0.29}^{+0.23}$  \\
\hline\noalign{\smallskip}
\multicolumn{11}{c}{\bf Stars not selected as Kepler targets}\\
\noalign{\smallskip}\hline
91178    &              &  9.10 &  1.79\,$\pm$\,0.75  & A2\,$\pm$\,2 &  9900\,$\pm$\,200  & 4.0\,$\pm$\,0.1 & 110 & 4306.463 & $-$11.8\,$\pm$\,1.4 & 1.96$_{-0.49}^{+0.32}$ \\     
93924    &              &  9.09 &  1.47\,$\pm$\,0.78  &	... 	      & 10400\,$\pm$\,300  & 4.2\,$\pm$\,0.2 & 150 & 4306.352 & $-$39.6\,$\pm$\,6.5 & ...~~~~~ \\
94137    &              &  9.83 &  2.17\,$\pm$\,0.88  &	... 	      &  7200\,$\pm$\,200  & 4.0\,$\pm$\,0.4 &  50 & 4311.580 & $-$28.0\,$\pm$\,1.2 & 1.37$_{-0.47}^{+0.32}$ \\
95069    &              &  9.05 &  2.70\,$\pm$\,1.09  & A1\,$\pm$\,2 & 12300\,$\pm$\,700  & 4.0\,$\pm$\,0.2 & 200 & 4307.372 & $-$13.8\,$\pm$\,1.7 & 1.77$_{-0.47}^{+0.31}$ \\
95506    &              &  9.15 &  3.66\,$\pm$\,0.69  & A1\,$\pm$\,2 & 10200\,$\pm$\,400  & 4.0\,$\pm$\,0.3 & 200 & 4309.345 & $-$22.8\,$\pm$\,4.6 & 1.26$_{-0.20}^{+0.17}$ \\
96061$^3$&              &  9.64 &  1.75\,$\pm$\,6.82  & A3\,$\pm$\,3 & 11300\,$\pm$\,500  & 4.2\,$\pm$\,0.2 &  70 & 4285.370 & $-$22.0\,$\pm$\,2.3 & ...~~~~~ \\
~~~$\prime\prime$~~~ &~~~~~~~~&~~~$\prime\prime$~~~ &~~~$\prime\prime$~~~& &~~~$\prime\prime$~~~ &~~~$\prime\prime$~~~ &$\prime\prime$ & 5108.346  &  $-$20.5\,$\pm$\,1.6 & \\
96066    &              &  8.80 &  4.04\,$\pm$\,2.13  & A1\,$\pm$\,3 & 10400\,$\pm$\,300  & 4.1\,$\pm$\,0.2 &  50 & 5103.450 &  $-$28.5\,$\pm$\,2.2   & ...~~~~~ \\
96343    &              &  9.70 &  3.80\,$\pm$\,0.91  & B9\,$\pm$\,3 & 11700\,$\pm$\,600  & 4.1\,$\pm$\,0.2 & 160 & 4311.365 &  $-$4.5\,$\pm$\,3.1 & 1.14$_{-0.26}^{+0.21}$ \\
96776    &              &  9.52 &  1.60\,$\pm$\,0.74  &	... 	      &  7600\,$\pm$\,200  & 4.5\,$\pm$\,0.5 & 200 & 4310.541 &  $-$7.7\,$\pm$\,3.6 & 1.80$_{-0.56}^{+0.35}$ \\
98486    &              &  9.59 &  1.20\,$\pm$\,0.90  &	... 	      & 10300\,$\pm$\,400  & 4.0\,$\pm$\,0.2 &  70 & 4308.409 &  $-$23.8\,$\pm$\,3.2 &  ...~~~~~ \\
98551\,A &              & 10.36 &  2.17\,$\pm$\,1.18  & ... 	      & ...~~~~		 & ...  	   &  70 & 4313.538 & $-$23.4\,$\pm$\,3.3    &  ...~~~~~ \\
98551\,B &              &~~~$\prime\prime$~~~ &~~~$\prime\prime$~~~& ... & ...~~~~ & ... & ...    & 4313.538 &    44.6\,$\pm$\,2.0 &  \\
98814    &              & 10.10 &  1.45\,$\pm$\,0.91  & B6\,$\pm$\,3  & 13900\,$\pm$\,1000 & 3.9\,$\pm$\,0.3 &  40 & 4313.372 & $-$40.7\,$\pm$\,3.5 &  ...~~~~~ \\    
\hline
\end{tabular}\\
\begin{flushleft}
 $^{\rm a}$ $V$ magnitude from the Hipparcos main catalogue \citep{esa97}.\\
 $^{\rm b}$ Parallax from \citet{vanLeeuwen}.\\
 $^{\rm c}$ Luminosity in solar units computed only for those stars with relative parallax errors $\le$\,50$\%$.\\
 $^1$ He-weak. Gravity in disagreement with the position on the HR diagram.\\ 
 $^2$ HgMn star.\\
 $^3$ Based on the proper motions \citep{esa97} and radial velocities, we suppose that HIP\,96061 is likely a physical binary with HIP~96066 with a separation of $\approx 11\arcsec$.5. 

\end{flushleft}
\end{center} 
\end{minipage}
\end{table*}

\begin{figure}
\includegraphics[width=8.45cm]{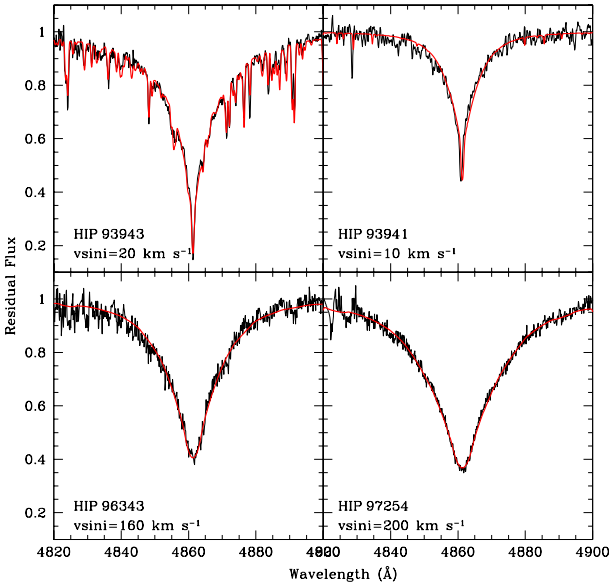}
\caption[]{Example of fitting the synthetic to the observed H$\beta$ lines for four stars from
our sample having different rotational velocities and/or different effective temperatures.}
\label{hbeta}
\end{figure}

\section{Determination of the atmospheric parameters}

We determined the effective temperature, $T_{\rm eff}$, and surface gravity, $\log g$,
of our program stars by minimizing the difference between the observed and the synthetic 
H$\beta$ profiles. For the goodness-of-fit parameter, we used $\chi^2$ defined as\\

$\displaystyle \chi^2 = \frac{1}{N} \sum \bigg(\frac{I_{\rm obs} - I_{\rm th}}{\delta I_{\rm obs}}\bigg)^2$,

\bigskip

\noindent
where $N$ is the total number of points, $I_{\rm obs}$ and $I_{\rm th}$ 
are the intensities of the observed and computed profiles, respectively, 
and $\delta I_{\rm obs}$ is the photon noise. Errors were estimated to be the variation 
in the parameters that increases the $\chi^2$ by unity.
As starting values of T$_{\rm eff}$ and $\log g$, we determined the effective 
temperature and gravity from our Str\"omgren photometry according to the 
grid of \citet{md85}. The photometric colors were dereddened with the 
\citet{Moon85} algorithm. For stars not photometrically observed by us, we used 
the values quoted by \citet{molenda2006}. We also proceeded with
the determination of rotational velocities by matching Mg{\sc ii}~$\lambda$4481~{\AA}
profile with a synthetic one. The results of our calculations are reported in Cols. 7, 8 and 9 of Table~\ref{param}. 

In Fig.~\ref{hbeta}, we show four examples of fitting the synthetic H$\beta$ profile
to the observed line in stars with different rotational velocities. The
theoretical profiles were computed with SYNTHE \citep{kur81} on the basis 
of ATLAS9 \citep{kur93} atmosphere models. All the models were computed using
solar opacity distribution function (ODF) and microturbulence velocity $\xi$\,=\,2~km~s$^{-1}$. 
All main sequence or slightly evolved stars ($\log g \geq $ 3.0) have microturbulences 
between 1.5 and 3.5 km/s (e.g., \citealt{gray01}, and references therein). Neglecting 
this small dependence on $\xi$, we can introduce a systematic error 
not larger than 0.1 dex into the abundance determination.

\begin{figure}
\includegraphics[width=8.45cm]{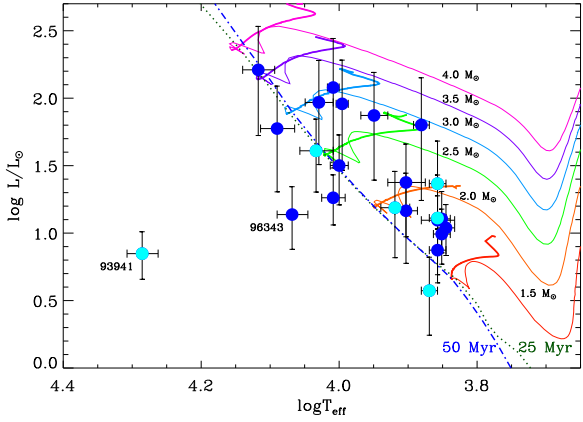}
\caption[]{HR diagram of the stars with relative parallax error $\le$\,50$\%$.
Light-grey (cyan) dots represent the targets for which a detailed abundance analysis has been 
performed. Evolutionary tracks for stellar masses ranging from 1.5 to 4.0 solar
masses \citep{siess00} have been also displayed (thick lines have been used for the post-main
sequence portion of each track). The isochrones at 25 and 50 Myrs are plotted 
with dotted and dash-dotted lines, respectively.}
\label{hr}
\end{figure}

In Fig.~\ref{hr}, we placed the program stars with parallax errors smaller than 50\% on the 
$\log T_{\rm eff}$ -- $\log L/L_{\odot}$ diagram that we constructed using effective temperatures 
derived in this paper and the luminosities calculated from both the $V$ magnitudes and the parallaxes 
reported by \citet{vanLeeuwen} and listed in Table~\ref{param}.
We evaluated the interstellar extinction $A_{\rm V}$ on the basis the star's distance, assuming a 
mean extinction of 1.7 mag/kpc on the Galactic plane ($|b| < 5\degr$) and 0.7 mag/kpc out of the plane. 
The de-reddened magnitude was converted into absolute magnitude $M_{\rm V}$ with the parallax and 
subsequently converted into bolometric magnitude by using the bolometric correction tabulated by 
\citet{Flower96} as a function of the effective temperature. The bolometric magnitude of the Sun, 
$M_{\rm bol} = 4.64$ \citep{Cox00}, was used to express the stellar luminosity in solar units.
The uncertainty in the stellar luminosity accounts for the parallax error. 

As can be seen in the figure, most stars are located on the main sequence or slightly above it.
Only two stars fall below the main sequence, accounting also for the errors in $T_{\rm eff}$ and
luminosity, namely HIP\,96343 and HIP\,93941. The first lies near the position of metal poor stars, 
but this is incosistent with its iron abundance ([Fe/H]$\sim0.5$, see Table\,\ref{abundFe}).
This inconsistency could be attributed to the companion star at 9$\arcsec$, which may have affected the 
observed spectrum leading to wrong stellar parameters. The discrepancy is far more evident for HIP\,93941, 
whose position in the $\log T_{\rm eff}$ -- $\log L/L_{\odot}$ diagram is inconsistent with
the star's surface gravity (see Table~\ref{param}). We discuss this star in more detail 
in Sect.~\ref{sec:93941}.

\subsection{Spectral Classification}

We performed a spectral classification of our program stars following the guidelines
of \citet{Herna04}, who apply their scheme to low-resolution spectra. For
this purpose, we degraded the resolution of the observed spectra from R\,=\,21\,000 to
R\,=\,3\,000 by convolving the spectra with a Gaussian kernel of the appropriate width. We 
then, measured the equivalent widths ($EW$), of H$\gamma$, 
He{\sc i}$\lambda$4471, He{\sc i}$\lambda$5876, Mg{\sc ii}$\lambda$4481, and other 
photospheric lines useful to the spectral classification of hot stars.
Finally, we assigned spectral types to our program stars using the
relation between $EW$ and the spectral type tabulated by \citet{Herna04}. The 
uncertainty in our determinations, derived from the agreement between different diagnostics,
is typically one spectral subclass but can reach two subclasses for rapid rotators
and/or stars that have spectrograms of low signal-to-noise ratio. The spectral 
types assigned by us to 17 of our program stars are listed in the fifth column of 
Table~\ref{param}. The remaining 19 stars had too high $v\sin i$, too low SNR in
their spectrograms, or too low temperature for measuring helium and Mg{\sc ii} lines, and
remained unclassified.

\section{Metal abundances}

We determined stellar abundances of our program stars by computing synthetic spectra that
reproduce the observed ones. We therefore, divided the measured spectrograms
into several intervals, each 25 {\AA} wide, and derived the abundances in each interval
by performing a $\chi^2$ minimization of the difference between the observed and synthetic spectrum.
We adopted lists of spectral lines and atomic parameters from \citet{castelli04},
who updated the parameters listed originally by \citet{kur95}.

We computed the abundances relative to the solar standard
values given by \citet{asplund05}. For each element, we calculated the uncertainty
in the abundance to be the standard deviation of the mean obtained from individual 
determinations in each interval of the analyzed spectrum. For elements whose lines
occurred in one or two intervals only, the error in the abundance ($\sim$\,0.1 dex)
was evaluated by varying the effective temperature and gravity 
within their uncertainties given in Table~\ref{param}, $[ T_{\rm eff}\,\pm\, \delta 
T_{\rm eff}]$ and $[\log g\,\pm\,\delta \log g]$, and computing the abundance for 
$T_{\rm eff}$ and $\log g$ values in these ranges.
In Table~\ref{abund}, we list the abundances derived with the method described above
and applied to 12 stars from Table~\ref{param} for which $v\sin i$ does not exceed 
90 km s$^{-1}$. 

\begin{table*}
\caption{Chemical abundances computed for all the low-rotating stars (v $\sin
  i <$ 90 km s$^{-1}$) of our sample. The abundances are given in terms of
the solar ones \citep{asplund05}.}
\scriptsize
\label{abund} 
\begin{center}            
\begin{tabular}{l r r r r r r r r r r r r}            
\hline            
\hline            
~Z~$\left[\epsilon (X) \right]$ & HIP\,93522 & HIP\,93941     & HIP\,93943         &  HIP\,94137         &    HIP\,94670     &    HIP\,95092      &    HIP\,96061     &    HIP\,96066     &    HIP\,96210      &    HIP\,96277       &  HIP\,96762         &  HIP\,98486  \\
\hline                                                                                                                                                                                                             
~~6~~$\left[ C  \right]$ & 0.2\,$\pm$\,0.2 & 0.4\,$\pm$\,0.3    & 0.1\,$\pm$\,0.3    & 0.2\,$\pm$\,0.4     & ---~~~~           & ---~~~~            & ---~~~~           & ---~~~~           & 1.1\,$\pm$\,0.1    & $-$0.3\,$\pm$\,0.4  &  0.6\,$\pm$\,0.1    &  ---~~~~  \\
~~7~~$\left[ N  \right]$ &    ---~~~~      & 1.5\,$\pm$\,0.1    & ---~~~~            & ---~~~~             & ---~~~~           & ---~~~~            & ---~~~~           & ---~~~~           & ---~~~~            & ---~~~~             & ---~~~~             &  ---~~~~  \\
~~8~~$\left[ O  \right]$ & 0.2\,$\pm$\,0.3 & 0.6\,$\pm$\,0.1    & 0.1\,$\pm$\,0.4    & 0.3\,$\pm$\,0.3     & 0.2\,$\pm$\,0.2   & ---~~~~            & ---~~~~           & ---~~~~           & 0.7\,$\pm$\,0.2    & ---~~~~             & ---~~~~             &  ---~~~~  \\
10~~$\left[ Ne \right]$ &    ---~~~~      & 0.8\,$\pm$\,0.3    & ---~~~~            &  ---~~~~            & ---~~~~           & ---~~~~            & ---~~~~           & ---~~~~           & ---~~~~            & ---~~~~             & ---~~~~             &  ---~~~~  \\
11~~$\left[ Na \right]$ & 0.3\,$\pm$\,0.3 &   ---~~~~          & 0.6\,$\pm$\,0.5    & 0.7\,$\pm$\,0.3     & ---~~~~           & ---~~~~            & ---~~~~           & ---~~~~           & ---~~~~            & ---~~~~             & ---~~~~             &  ---~~~~  \\
12~~$\left[ Mg \right]$ & 0.4\,$\pm$\,0.3 & $-$0.6\,$\pm$\,0.1 & 0.0\,$\pm$\,0.2    & $-$0.1\,$\pm$\,0.3  & 0.2\,$\pm$\,0.2   & 0.2\,$\pm$\,0.5    & 0.0\,$\pm$\,0.1   & 0.1\,$\pm$\,0.2   & $-$1.3\,$\pm$\,0.1 & ---~~~~             & ---~~~~             & $-$0.1\,$\pm$\,0.4 \\
13~~$\left[ Al \right]$ & 0.1\,$\pm$\,0.1 & 0.6\,$\pm$\,0.4    & ---~~~~            & ---~~~~             & ---~~~~           & ---~~~~            & ---~~~~           & ---~~~~           & ---~~~~            & 0.1\,$\pm$\,0.2     &  $-$0.1\,$\pm$\,0.4 &  ---~~~~   \\
14~~$\left[ Si \right]$ &    ---~~~~      & 2.0\,$\pm$\,0.2    & 0.0\,$\pm$\,0.4    & $-$0.3\,$\pm$\,0.3  & ---~~~~           & ---~~~~            & ---~~~~           & ---~~~~           & $-$0.4\,$\pm$\,0.1 & 0.1\,$\pm$\,0.4     & ---~~~~             &  ---~~~~  \\
16~~$\left[ S  \right]$ &                 & 0.3\,$\pm$\,0.4    & $-$0.1\,$\pm$\,0.4 & ---~~~~             & ---~~~~           & ---~~~~            & ---~~~~           & ---~~~~           & ---~~~~            & 0.3\,$\pm$\,0.1     & ---~~~~             &  ---~~~~  \\
20~~$\left[ Ca \right]$ & 0.5\,$\pm$\,0.5 & 1.3\,$\pm$\,0.x    & 0.0\,$\pm$\,0.2    & $-$0.2\,$\pm$\,0.4  & ---~~~~           & $-$0.2\,$\pm$\,0.1 & 0.5\,$\pm$\,0.1   & ---~~~~           & ---~~~~            & $-$0.3\,$\pm$\,0.3  &  0.6\,$\pm$\,0.2    & 0.0\,$\pm$\,0.2\\
21~~$\left[ Sc \right]$ & 0.6\,$\pm$\,0.6 & ---~~~~            & 0.5\,$\pm$\,0.5    & 0.2\,$\pm$\,0.2     & ---~~~~           & 0.5\,$\pm$\,0.2    & ---~~~~           & ---~~~~           & ---~~~~            & $-$0.2\,$\pm$\,0.5  &  0.7\,$\pm$\,0.1    &  ---~~~~  \\
22~~$\left[ Ti \right]$ & 0.3\,$\pm$\,0.2 & 0.9\,$\pm$\,0.1    & 0.3\,$\pm$\,0.4    & $-$0.1\,$\pm$\,0.3  & ---~~~~           & $-$1.0\,$\pm$\,0.1 & 1.0\,$\pm$\,0.4   & 0.2\,$\pm$\,0.1   & 0.8\,$\pm$\,0.4    & $-$0.4\,$\pm$\,0.2  &  0.6\,$\pm$\,0.3    & 0.5\,$\pm$\,0.1\\
23~~$\left[ V  \right]$ &    ---~~~~      & ---~~~~            & 0.3\,$\pm$\,0.5    & ---~~~~             & ---~~~~           & ---~~~~            & ---~~~~           & ---~~~~           & ---~~~~            & 0.8\,$\pm$\,0.4     & ---~~~~             &  ---~~~~  \\
24~~$\left[ Cr \right]$ & 0.4\,$\pm$\,0.2 & 1.6\,$\pm$\,0.5    & 0.0\,$\pm$\,0.2    & $-$0.5\,$\pm$\,0.1  & ---~~~~           & $-$0.3\,$\pm$\,0.4 & 0.3\,$\pm$\,0.1   & 0.0\,$\pm$\,0.1   & 0.9\,$\pm$\,0.2    & $-$0.9\,$\pm$\,0.2  &  0.8\,$\pm$\,0.4    & 0.3\,$\pm$\,0.1\\
25~~$\left[ Mn \right]$ & 0.6\,$\pm$\,0.4 & ---~~~~            & ---~~~~            & ---~~~~             & ---~~~~           & ---~~~~            & ---~~~~           & ---~~~~           & 2.2\,$\pm$\,0.2    & ---~~~~             & ---~~~~             &  ---~~~~  \\
26~~$\left[ Fe \right]$ & 0.4\,$\pm$\,0.1 & 1.2\,$\pm$\,0.4    & 0.0\,$\pm$\,0.1    & 0.1\,$\pm$\,0.3     & 0.1\,$\pm$\,0.3   & $-$0.5\,$\pm$\,0.5 & 0.6\,$\pm$\,0.2   & 0.2\,$\pm$\,0.1   & 0.2\,$\pm$\,0.4    & 0.0\,$\pm$\,0.5     &  $-$0.1\,$\pm$\,0.4 & 0.6\,$\pm$\,0.1\\
28~~$\left[ Ni \right]$ & 0.6\,$\pm$\,0.2 & ---~~~~            & $-$0.3\,$\pm$\,0.2 & ---~~~~             & ---~~~~           & ---~~~~            & ---~~~~           & ---~~~~           & 1.1\,$\pm$\,0.2    & $-$0.4\,$\pm$\,0.3  & ---~~~~             &  ---~~~~  \\
39~~$\left[ Y  \right]$ & 0.0\,$\pm$\,0.2 & ---~~~~            & 0.0\,$\pm$\,0.2    & 0.0\,$\pm$\,0.2     & ---~~~~           & ---~~~~            & ---~~~~           & ---~~~~           & ---~~~~            & 0.0\,$\pm$\,0.3     & ---~~~~             &  ---~~~~  \\
40~~$\left[ Zr \right]$ &   ---~~~~       & ---~~~~            & 0.0\,$\pm$\,0.2    &  ---~~~~            & ---~~~~           & ---~~~~            & ---~~~~           & ---~~~~           & ---~~~~            & ---~~~~             & ---~~~~             &  ---~~~~  \\
56~~$\left[ Ba \right]$ & 1.0\,$\pm$\,0.3 & ---~~~~            & 1.0\,$\pm$\,0.2    & 1.0\,$\pm$\,0.1     & ---~~~~           & ---~~~~            & ---~~~~           & ---~~~~           & ---~~~~            & 1.0\,$\pm$\,0.3     & ---~~~~             &  ---~~~~  \\
\hline
\end{tabular} 
\end{center} 
\end{table*}

\begin{figure*}
\includegraphics[width=14cm,angle=-90]{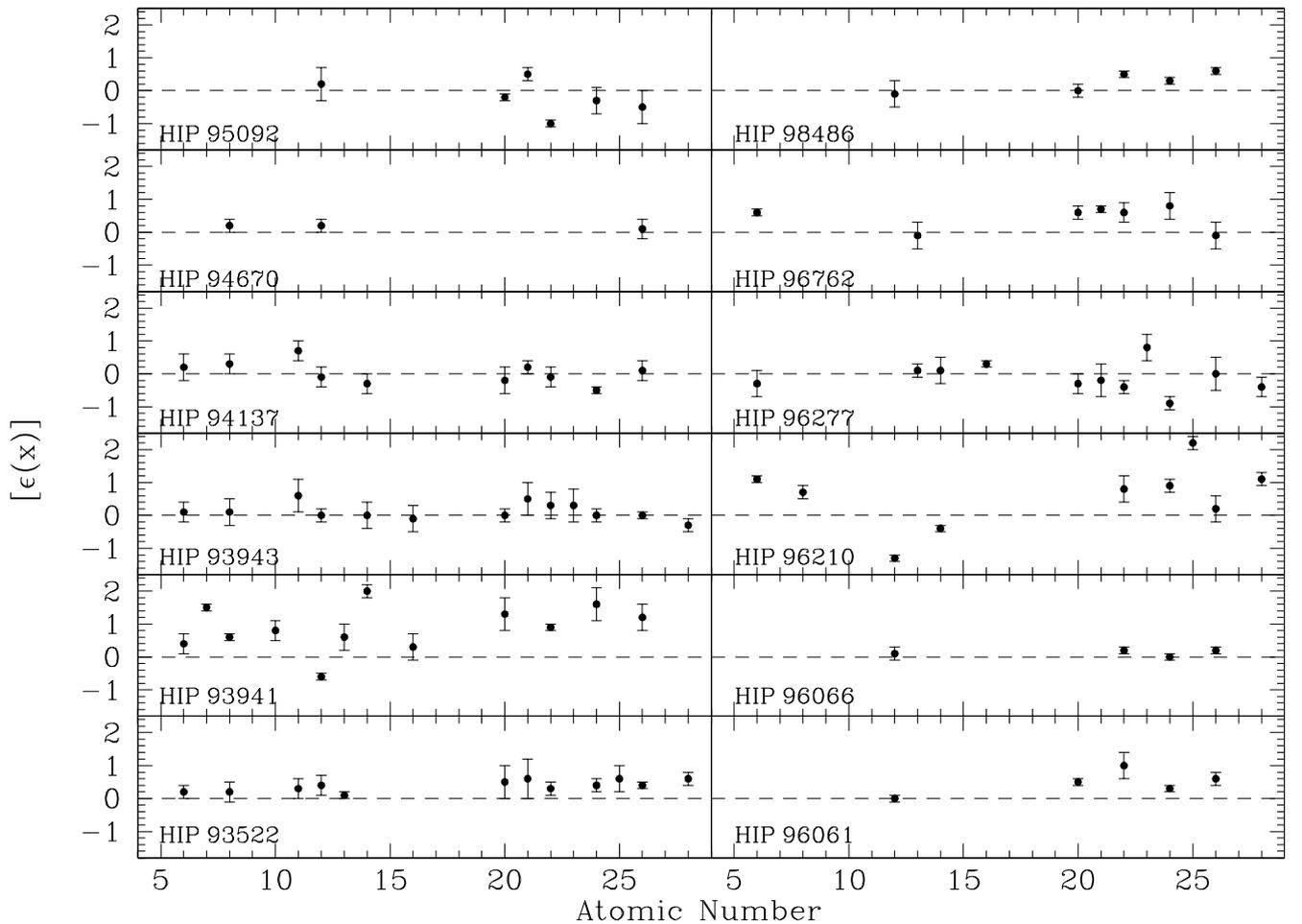}
\caption[]{Abundance patterns derived for 12 stars (see Table~\ref{abund}).
For the sake of clarity, we excluded from the plot all the chemical elements
with Z\,$>$\,28 (i.e., nickel).}
\label{pattern}
\end{figure*}

For stars with $v\sin i > 90$ km s$^{-1}$, the lines are too broad to attempt 
this kind of analysis and we derived only iron abundance from the equivalent 
widths of Fe{\sc ii} $\lambda \lambda$5018.44, 5316.615~{\AA}.
The latter were converted into abundances using WIDTH9 \citep{kur81} and the ATLAS9 
atmospheric models. The results obtained for rapid rotators are listed in Table~\ref{abundFe}.

For all stars apart from HIP\,93941, HIP\,95092, HIP\,96061, and HIP\,96210, which 
we discuss below, the metal abundances derived in this paper agree with the solar one
to within the error bars. We illustrate this in Fig.~\ref{pattern}, in which we plot
the abundance patterns of elements with the atomic number Z\,$\le$\,28, derived for the twelve 
stars listed in Table~\ref{abund}. The heavier elements are not shown for the sake of the
clarity in the diagrams.

\subsection{HIP\,93941}\label{sec:93941}
This star was classified to be spectral type B2 (with an uncertainty of two spectral 
sub-classes) for the first time in this paper. Our classification
is consistent with the star's ultraviolet magnitudes \citep[see][]{carnochan1983}
measured by the sky survey S2/68 telescope on-board the ESRO TD1 satellite \citep{Boksenberg1973}
and with its H$\beta$ photoelectric photometry by \citet{Deutschman1976}.

Our detailed abundance analysis indicates that HIP\,93941 has significant overabundances of 
silicon ($\approx$\,2~dex), chromium ($\approx$\,1.6~dex), and nitrogen ($\approx$\,1.5~dex), 
as well as moderate overabundances of calcium, titanium, and iron (all $\approx$\,1~dex) and an 
underabundance of magnesium. This specific pattern of element abundances resembles the 
abundances measured in the He-weak star $\alpha$ Scl \citep[see][]{lopez2001}. Taking into 
account its very low projected rotational velocity, $v\sin i$ = 10 km s$^{-1}$,
which is typical of this class of stars, we suspected that HIP\,93941 is a He-weak star. 
Thus, to confirm this kind of peculiarity, we attempted to derive the helium abundance
by performing a spectral synthesis of the He\textsc{I}\,4471\,\AA line. According to \citet{LeoLanza}, 
the behavior of this line is quite independent of the microturbulence, and NLTE effects are 
small for $T_{\rm eff}<20\,000$\,K. For HIP\,93941, we derive an helium abundance [He]=$-0.30\,\pm\,0.10$. 
The error was evaluated by computing the abundance for T$_{\rm eff}$ and $\log g$ within the intervals 
$[ T_{\rm eff}\,\pm\, \delta T_{\rm eff}]$ and $[\log g\,\pm\,\delta \log g]$.
This make HIP\,93941 a very interesting target for the asteroseismic part of the Kepler mission
because pulsations have not yet been detected in He-weak stars.

\subsection{HIP\,95092 and HIP\,96061}
For these main-sequence stars, normal abundances were derived. The only exception is titanium,
which is underabundant by $\approx$\,1~dex in HIP\,95092 and overabundant by $\approx$\,1~dex 
in HIP\,96061. Similar anomalies have also been observed in other normal A-type stars, i.e., 
Vega \citep{lemke1989} or $\beta$\,Pic \citep{Holweger1997}, so we can consider HIP\,95092 and 
HIP\,96061 as normal stars.

\subsection{HIP\,96210}
HIP\,96210 shows a general overabundance of carbon, oxygen, and all iron-peak elements with a 
particularly strong overabundance of manganese ($\approx$\,2.2~dex) and a strong underabundance 
of magnesium. With respect to the manganese abundance, and taking into account the values of $T_{\rm eff}$ 
and $\log g$ for this star, HIP\,96210 should be classified as a member of the group of 
HgMn chemically peculiar stars \citep[see, e.g.,][]{Jaschek1987}. We note, however, that 
this classification needs to be confirmed in additional observations because the spectral range 
of our data do not cover the blue region of the spectrum around the Hg{\sc i} 
$\lambda$3984 {\AA}, which is an important diagnostic for this kind of peculiarity.

\begin{table}
\caption{Iron abundances computed for all the rapid-rotating stars. 
Typical errors are on the order of 0.2 dex} 
\label{abundFe} 
\begin{center}            
\begin{tabular}{l r}                 
\hline            
\hline            
HIP  & $\left[\epsilon (Fe) \right]$ \\
\hline
91178 & 0.1~~ \\
92247 & 0.4~~ \\
92259 & 0.2~~ \\
93924 & 0.3~~ \\
94809 & 0.4~~ \\
95174 & $-$0.1~~ \\
95495 & 0.4~~ \\
95506 & 0.5~~ \\
96343 & 0.6~~ \\
96776 & 0.7~~ \\
97254 & 0.1~~ \\
97486 & 0.3~~ \\
97582 & 0.2~~ \\
97724 & $-$0.2~~ \\
98160 & $-$0.1~~ \\
98814 & 0.2~~ \\
\hline
\end{tabular} 
\end{center} 
\end{table}

\section{Stars variable in radial velocity}

Around 50\%  of stars of all spectral types are binaries \citep[see][]{Petrie1960}. This also 
concerns stars selected as Kepler asteroseismic targets. Since an additional component influences 
the star's magnitude and colors, the contribution from the secondary star has to be computed and removed 
from the Kepler data before the asteroseismic analysis can begin.

Below, we discuss eight stars that we find to be variable in radial velocity either in our observations
or by comparing our results with those published in the literature. In all cases, additional observations 
are required to study these systems in more detail, and to confirm that the other stars are singles or to 
eventually detect new spectroscopic binaries among them.

\subsection{Double-lined spectroscopic binaries}\label{sec:SB2}

\medskip
\noindent
{\it HIP\,96299}
\smallskip

\noindent
This star was discovered to be an eclipsing binary with a period of 10.0486 days by 
\citet{Hartman2004}. We found that it also appears as
a double-lined spectroscopic binary. This makes HIP\,96299 a very interesting target 
for the Kepler asteroseismic study since for SB2 eclipsing binaries it is possible to derive 
accurate values for the mass of the two components, which is a basic ingredient for constraining 
evolutionary and asteroseismic models. 
As we acquired only one spectrogram of this star, further observations are needed to obtain 
its radial-velocity curve and determine its orbital solution.

\medskip
\noindent
{\it HIP\,98551}
\smallskip

\noindent
This star, which was not selected as a Kepler asteroseismic target, is the second double-lined 
spectroscopic binary discovered by ourselves. As in the previous case, we have only one spectrogram of 
this target. HIP\,98551 is not known to exhibit eclipses.

\subsection{Single-lined spectroscopic binaries}

For single-lined binary stars, we cannot derive the masses of the components in the
above-mentioned way. We need instead to calculate the systems' mass function: this 
can be used to estimate the magnitude and color indices of the secondary component of the 
system and allow us to calculate the duplicity corrections for the primary. In Table~\ref{rv}, 
we list five stars that we find to be variable in $RV$. The table also includes the two SB2 systems 
discussed in Sect.\ref{sec:SB2}, and an already known SB1 system, namely HIP\,98814.

\medskip
\noindent
{\it HIP\,92637}
\smallskip

\noindent
This star was observed spectroscopically by \citet{Dworetsky1982}, who used the X-spectrograph mounted
on the 1.5 telescope at Mount Wilson during May-August 1975 with the aim of classifying to the MK system 
the `ultraviolet objects' from the S2/68 experiment. \citet{Dworetsky1982} classify HIP\,92637 as B4V:, 
which is in a rough agreement with our classification, and measure the star's radial velocity, 
to be 27$\pm$36 km/s. The measurement of $RV$ by \citet{Dworetsky1982} is very inaccurate but still 
differs significantly from the $RV$ measured in our spectra. We then classify HIP\,92637 as a suspected SB1,
which should be targeted by future observations.

\begin{table}
\caption{Spectroscopic binaries and stars with possible $RV$ variations.}
\label{rv} 
\begin{center}            
\begin{tabular}{r r r c l}                  
\hline            
\hline            
HIP      & $RV$ [km/s]& $RV$ [km/s]& ref. & var \\
         & this paper & literature &      & \\
\hline
92637    &  $-$22.72\,$\pm$\,2.58 &    27\,$\pm$\,36~~ & (1) & SB1?\\
96277    &  $-$0.52\,$\pm$\,1.16$^{\rm a}$  &      ...~~~~~      &...  & ~~?\\
96299\,A &  $-$7.77\,$\pm$\,1.22  &      ...~~~~~      &...  & SB2\\ 
96299\,B &     45.13\,$\pm$\,1.26 &      ...~~~~~      &...  & ...\\
96762    &     16.51\,$\pm$\,1.10 &    24\,$\pm$\,6.2~ & (2) & SB1?\\
97254    &   $-$4.28\,$\pm$\,6.19 & $-$23\,$\pm$\,1.5~ & (2) & SB1\\
97724    &  $-$14.76\,$\pm$\,4.05 & $-$31\,$\pm$\,3.4~ & (2) & SB1\\
98551\,A &  $-$23.44\,$\pm$\,3.29 &      ...~~~~~      &...  & SB2\\
98551\,B &     44.64\,$\pm$\,1.97 &      ...~~~~~      &...  & ...\\
98814    &  $-$40.70\,$\pm$\,3.51 & $\gamma = 8$~~     & (4) & SB1\\
\hline
\multicolumn{5}{l}{
\begin{minipage}{6.6cm}
$^{\rm a}$ Mean of the three $RV$ values reported in Table~\ref{param}.\\
(1) \citet{Dworetsky1982}, \\
(2) \citet{Fehrenbach1990},\\
(3) \citet{Fehrenbach1997},\\
(4) \citet{Struve1946}\\
\end{minipage}
}
\end{tabular}
\end{center} 
\end{table}

\medskip
\noindent
{\it HIP\,96277}
\smallskip

\noindent
The $RV$ of this star shows a peak-to-peak scatter in our three spectra of the order of three times the 
average uncertainty in our measurements. Therefore, we classify it as a possible SB1 and 
note that additional observations are needed to confirm this finding.

\medskip
\noindent
{\it HIP\,96762}
\smallskip 

\noindent
This star was observed by \citet{Fehrenbach1990} with the objective prism at the Observatorie de
Haute-Provence. The radial velocity measured by these authors, 24\,$\pm$\,6 km\,s$^{-1}$,
notwithstanding the relatively large uncertainty, is  higher than the $RV$ reported in 
the present paper (16.5\,$\pm$\,1.1 km\,s$^{-1}$) by 1.2 $\sigma_{RV}$. As such, this is another 
suspected SB1 system.

\medskip
\noindent
{\it HIP\,97254}
\smallskip

\noindent
This is another star observed by \citet{Fehrenbach1990}. The significant difference between the $RV$ 
measured by these authors and that reported in this paper, $\Delta RV\simeq 13$\,km\,s$^{-1}$ 
(i.e., more than 3 $\sigma_{RV}$), allows us to conclude that HIP\,97254 is a new SB1 system.

\medskip
\noindent
{\it HIP\,97724}
\smallskip

\noindent
For this star also the radial velocity measured by \citet{Fehrenbach1990} differs significantly
from the value measured by ourselves. Therefore, we classify HIP\,97724 as the second new SB1 star 
in our sample.

\medskip
\noindent
{\it HIP\,98814}
\smallskip

\noindent
This star was discovered to be an eclipsing binary of the Algol type by \citet{Ceraski1904}.
The spectroscopic study of \citet{Struve1946} showed that HIP\,98814 is a single-lined spectroscopic 
binary with a circular orbit of semi-amplitude {\it K}\,=\,68\,km~s$^{-1}$ and an orbital period 
of 3.3177 days. Our single data-point, $RV$\,=\,$-$40.70$\pm$3.51, phased with the orbital 
period of 3.3178\,d determined by \citet{Hartman2004} from photometric observations, follows the 
$RV$ curve of \citet{Struve1946}, which has a rather large scatter being based on old plate spectra.

We note that the $RV$ curve obtained by \citet{Struve1946} remains the only determination 
available to date for this star. Two subsequent spectrograms acquired by \citet{Etzel1993} at the Mount 
Laguna Observatory (San Diego, California, USA) were used by the authors to derive the star's projected 
rotational velocity, of 41$\pm$7 km/s, which is in a very good agreement with the value reported in the
present paper. \citet{Etzel1993} do not provide the $RV$ for HIP\,98814 obtained from their data.

\section{Conclusions}

We have discussed the atmospheric and kinematical properties of 23 Kepler asteroseismic targets
and an additional 13 stars falling in the Kepler field of view that had not been selected as mission 
targets. We measured the stars' radial velocity and the projected 
rotational velocity, and derived the effective temperature, the surface gravity, and the abundances of 
up to 20 different species. In this last task, the number of analyzed 
species depended on the SNR of the individual spectrograms and the rotational velocity of the target. 

We identified two double-lined spectroscopic binaries, HIP\,96299 and HIP\,98551,
two single-lined spectroscopic binaries, HIP\,97254 and HIP\,97724, two suspected SB1 
systems, HIP\,92637 and HIP\,96762, and one star, HIP\,96277, whose tentative variations in $RV$
have to be confirmed with additional observations. More spectra are needed to measure the 
radial-velocity curves of these stars and compute the systems' orbital solutions, as well as attempt 
to detect secondary components in high-resolution spectrograms of the SB1 stars.
SB2 systems, especially those that are also eclipsing binaries, are important asteroseismic targets 
for Kepler since for them it is possible to compute the masses of the components, which places strong 
constraints on the asteroseismic models that fit the observed frequency spectrum.

For 17 stars in our sample, we computed a MK spectral classifications and for 28 stars we performed 
an abundance analysis. We also discovered two stars that are chemically peculiar, 
namely, HIP\,96210, which we classified as a suspected HgMn star, and HIP\,93941, classified as He-weak. 
A full, detailed analysis of the abundances of the Kepler targets will 
require acquisition of high-resolution, spectrograms of high signal-to-noise ratio covering a 
wide part of the optical domain, which is planned for the most interesting Kepler targets. 
However, such high-quality data could be probably obtained only for a quite small fraction
of all the stars observed by Kepler, due to their faintness and the availability of  
high-resolution echelle spectrographs at large-aperture telescopes in the northern hemisphere.

Therefore, we emphasize the importance of low- and medium-resolution spectroscopic and multicolor 
photometric observations of Kepler targets, which can be acquired for a high number of stars
with medium-sized telescopes and will allow us to maximize the scientific output of the mission.

Finally, we draw attention to five Kepler targets, HIP\,93522, 93941, 93943, 96210, and 96762, 
which are of very low projected rotational velocity ($v\sin i < 20$ km\,s$^{-1}$). Were they to be
true slowly-rotating stars (not nearly pole-on fast rotators), they would be very promising 
targets for asteroseismology since an unambiguous interpretation of the observed frequency spectrum 
would then be possible. We note that HIP\,93941, 96210 and 96762 are B-type stars, which makes 
them even more interesting because so slowly rotating stars of this spectral class are rare.

\begin{acknowledgements}
      JM\.Z acknowledges the MNiSW grant N203 014 31/2650 and thanks the Astronomical 
      Institute of the University of Wroc\l aw for financial support. We made use 
      of the SIMBAD database, operated at CDS, Strasbourg, France, and the NASA's 
      Astrophysics Data System.
\end{acknowledgements}

\end{document}